\documentclass[twocolumn,showpacs,preprintnumbers,amsmath,amssymb]{revtex4}
\usepackage{graphicx}

\begin{document}
\title{Exchange interaction between the triplet exciton and the localized spin in copper-phthalocyanine}
\author{Wei Wu}\email{wei.wu@ucl.ac.uk}
\affiliation{London Centre for Nanotechnology, University College London, Gower Street, London, WC1E 6BT, United Kingdom}

\begin{abstract}
Triplet excitonic state in the organic molecule may arise from a singlet excitation and the following inter-system crossing. Especially for a spin-bearing molecule, an exchange interaction between the triplet exciton and the original spin on the molecule can be expected. In this paper, such exchange interaction in copper-phthalocyanine (CuPc, spin-$\frac{1}{2}$) was investigated from first-principles by using density-functional theory within a variety of approximations to the exchange correlation, ranging from local-density approximation to long-range corrected hybrid-exchange functional. The magnitude of the computed exchange interaction is in the order of meV with the minimum value (1.5 meV, ferromagnetic) given by the long-range corrected hybrid-exchange functional CAM-B3LYP. This exchange interaction can therefore give rise to a spin coherence with an oscillation period in the order of picoseconds, which is much shorter than the triplet lifetime in CuPc (typically tens of nanoseconds). This implies that it might be possible to manipulate the localised spin on Cu experimentally using optical excitation and inter-system crossing well before the triplet state disappears. 
\end{abstract}


\maketitle
\section{Introduction}

The metal-phthalocyanine (MPc), in which a metallic ion is centred at a conjugated Pc ring, has attracted much attention recently due to its fascinating physical properties  \cite{torre2007, heutz2007,wang2010, wu2008, wu2011, chen2008,  wu2013a, wu2013b, wu2013c, wu2013d, marc2012}. Especially, the spin-bearing MPcs are of great interest. They can have very long spin-lattice relaxation times \cite{marc2012}, which may pave the way towards robust electron-spin qubit. Moreover, as organic semiconductors, due to their plasticity, they can be prepared in different forms such as bulk crystalline, thin films \cite{heutz2007,wu2013d}, and nanowires \cite{wang2010}. In conjugation with their magnetic tunability realised by the state-of-the-art molecular growth techniques \cite{heutz2007, wu2013d}, these advanced materials properties are highly promising in the applications including quantum information processing (QIP) \cite{marc2012} and organic spintronics \cite{natreview}. 

Copper-phthalocyanine (CuPc) \cite{heutz2007, wu2008, wang2010, wu2011} has received the most extensive experimental and theoretical studies as compared to other MPcs. 
Nevertheless, the recent experimental and theoretical studies have demonstrated that the $\alpha$-phase cobalt-phthalocyanine (CoPc) has much stronger exchange interaction than CuPc and raises its transition temperature above the boiling point of liquid nitrogen \cite{wu2013a, wu2013d}. This is probably due to the out-of-plane orientation of $d_{z^2}$-orbital in CoPc (in contrast to the in-plane orbital $d_{x^2-y^2}$ in CuPc), which might ease the migration of electron between molecules. These studies have illustrated the importance of choosing the appropriate metallic ion for the manipulation of $d$-orbital occupations and hence the magnetic properties. However, for practical application, it would be very desirable to manipulate the exchange interaction by external stimulus such as optical excitation. 

The optically excited state in molecule could not only facilitate the control of interaction in QIP \cite{timco2009} but also play an important role in organic spintronics  \cite{miller2000, xiong2004}. Control of the strongest interaction between electron spins, exchange interaction, so as to manipulate the entanglement, is an important aspect in quantum computation \cite{kane}. By controlling the exchange interaction via optical excitation, the additional electrodes that are supposed to control exchange interaction \cite{kane} but would induce significant decoherence~\cite{ams} can be avoided. In addition, excitons in organic materials could interact with a polaron \cite{desai2007, wagemans2010}, which can influence the spin transport and hence the magneto-resistance. Due to these important aspects, it is necessary to understand the interaction between spin and exciton in spin-bearing molecules.

The Pc ring has a singlet ground state, on which a triplet state can be induced by the inter-system crossing that follows the optical excitation of singlet state~\cite{porphyrinbooks,sato2007}. Therefore, the spin-bearing MPc, having a broad optical absorption spectrum \cite{wang2010}, is an ideal platform for studying optically induced exchange interaction. The experiments on the triplet state of CuPc have indicated its lifetime $T$ is very short ($\sim 35 \ ns$ in Ref.\cite{mcvie1978} and $< 50 \ ns$ in Ref.\cite{zhang1994} respectively) as compared to molecules such as metal-free phthalocyanine ($T\sim\mu s$) \cite{mcvie1978}. Therefore, another crucial question, which is related to QIP, is whether the spin coherence induced by the exchange interaction ($J$) can survive such fast relaxation of triplet state, i.e., according to Heisenberg's uncertainty principle, a comparison between $J$ and $\frac{\hbar}{T}$ is needed. Previously, the electronic structures of the ground and excited states of CuPc have been studied \cite{rosa1994, marom2008,vas2009} within density-functional theory (DFT). The electrostatic interactions between organic ligand and metallic ions for a variety of MPc including CuPc have been studied by using local-density approximation (LDA) within DFT in Ref.\cite{rosa1994}. The ground-state electronic structure of CuPc has been studied by using a variety of functionals within DFT in Ref.\cite{marom2008,vas2009}. However, the aspect mentioned above, especially the exchange interaction between triplet exciton and Cu spin, is yet to be investigated theoretically and experimentally. In addition, the dependence of the computed exchange interaction on the functional is of great interest for the wider research communities of DFT and organic magnetism as well. In this paper, a theoretical study of the exchange coupling and electronic structure in CuPc with a triplet on the Pc ring, was carried out within DFT by using a variety of approximations to the exchange-correlation functional ranging from LDA to long-range corrected hybrid-exchange functional. A strong exchange interaction has been predicted from first-principles and the nature of the triplet exciton has been investigated carefully. The calculation results relying on different approximations to the exchange-correlation functionals have been compared. More importantly, the magnitude of the computed exchange interaction could result in a spin coherent oscillation with a period in the order of picoseconds, which could potentially survive the fast relaxation of triplet in CuPc. The remaining content falls into three parts. In \S{\ref{compmethods}}, a brief introduction to the computational tools used is given. In \S\ref{results}, the calculated electronic structure and exchange interaction between the triplet and the localised spin are reported and the possible spin manipulation mechanism is discussed. Finally, in \S\ref{conclusions}, some more general conclusions are drawn.

\section{Computational details}\label{compmethods}





Calculations for an isolated CuPc molecule have been carried out using DFT and Dunning's correlation-consistent polarized valence double-zeta (cc-pVDZ) basis set \cite{dunning} implemented in the Gaussian 09 code \cite{gaussian09}. The self-consistent field (SCF) procedure is converged to a tolerance of $10^{-6}$ a.u. ($\sim 0.03$ meV). 

A number of approximations to the exchange-correlation functionals, starting from the simple LDA to the more sophisticated long-range corrected hybrid-exchange functional were used to describe the electronic exchange and correlation. Local spin-density approximation (LSDA) to the correlation functional, which was developed by Vosko, Wilk, and Nusair in the 1980s \cite{lsda1}, together with Slater's exchange functional \cite{lsda2}, was employed here. The generalised gradient approximations (GGA) used here include those invented by Perdew and Wang (PW91) \cite{pw91}, Perdew, Burke, Ernzerhof (PBE) \cite{pbe}, the further developed formalism for PBE (PBEh1PBE) \cite{pbeh1pbe}, and BLYP functional \cite{blyp1, blyp2}.  The hybrid-exchange functionals, O3LYP \cite{o3lyp}, B3LYP \cite{b3lyp}, and HSE06 \cite{hse06} have been chosen for the calculations. Recently the long-range corrected hybrid-exchange functional, CAM-B3LYP \cite{camb}, has been developed to improve B3LYP functional by mixing different amount of exact exchange in the long range as compared to that in the short range. LSDA is the simplest approximation for the exchange-correlation functional used in this paper. GGA approximations including PW91 \cite{pw91}, PBE \cite{pbe}, PBEh1PBE \cite{pbeh1pbe} and BLYP \cite{blyp1, blyp2} are slightly more sophisticated than LSDA by including the gradient of electron density as additional variable of the functional. In sharp contrast to the LDA and GGA, the hybrid-exchange functionals including O3LYP, B3LYP, and HSE06 can eliminate the self-interaction error and balance the tendencies to delocalize and localize Kohn-Sham orbitals by mixing Fock exchange with that from a generalized gradient approximation (GGA) exchange functional \cite{b3lyp}. The hybrid-exchange functionals have previously been shown to provide a good description of the localised spins and magnetic properties \cite{illas00, muscat2001, wu2008, wu2011,wu2013a}. Especially for CuPc (spin-$\frac{1}{2}$), it is crucial to treat the localised $d$-electron and delocalized $p$-electron wave functions on the same foot. As shown later on, within the broken-symmetry methods \cite{noodleman}, the choice of the functional indeed leads to a significant difference for the exchange interaction. 

The broken-symmetry method \cite{noodleman} was used to localize anti-aligned spins on the molecule. The Heisenberg spin Hamiltonian used to describe the interaction between spins reads \cite{heisenberg} 
\begin{equation}\label{eq:spinh}
\hat{H}=J{\hat{\vec{S}}\cdot\hat{\vec{s}}},
\end{equation}, where $\hat{\vec{S}}$ and $\hat{\vec{s}}$ represent spin-1 and spin-$\frac{1}{2}$ operators, respectively. The exchange constant $J$ was estimated from the electronic structure calculations as,
\begin{equation}\label{eq:DeltaE}
J=E_{\mathrm{FM}}-E_{\mathrm{AFM}},
\end{equation}
where $E_{\mathrm{AFM}}$ and $E_{\mathrm{FM}}$ are the total energies of the spin configurations that are anti-ferromagnetic (AFM, spins anti-aligned) and ferromagnetic (FM, spins aligned), respectively. When forming the AFM configuration, the related molecular orbitals were switched by using the keyword GUESS=ALTER in the Gaussian 09 code to initialise the desired quantum state. Supposing one starts from the ground state of CuPc (spin-$\frac{1}{2}$), the spin-up $d_{x^2-y^2}$-derived orbital need to be swapped with the lowest-unoccupied molecular orbital (LUMO), meanwhile, the spin-down unoccupied $d_{x^2-y^2}$-derived orbital need to be swapped with the highest-occupied molecular orbital (HOMO). The self-consistent process would then facilitate the formation of the AFM (broken-symmetry) state of the triplet exciton on the Pc ring and the Cu spin provided an appropriate functional is used.
\section{Results and discussion}\label{results}
\subsection{Electronic structure}
\begin{figure*}[htbp]
\centering
\includegraphics[scale=0.875,clip=true]{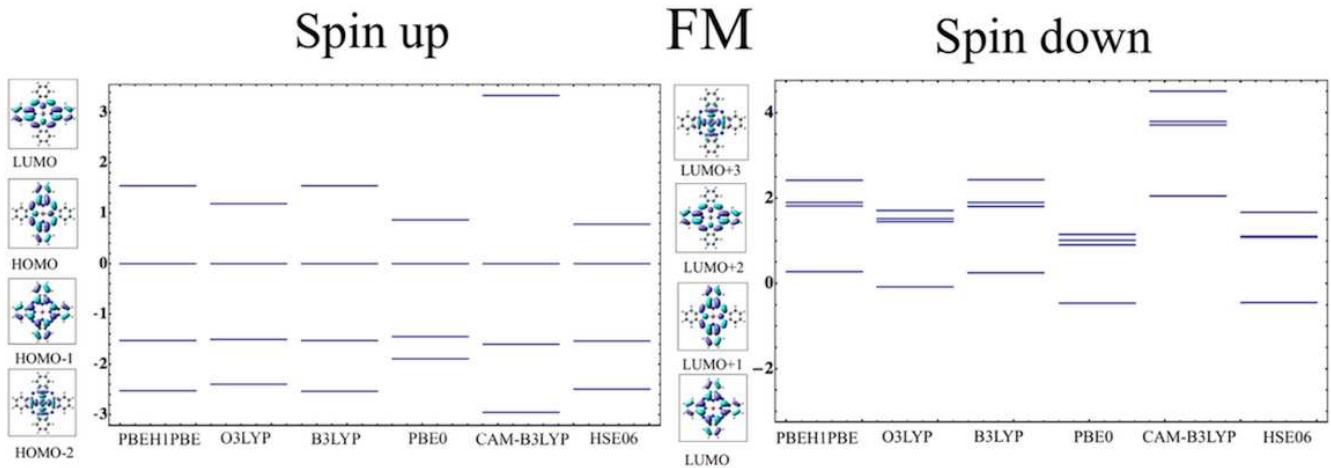}
\caption{(Color online.) The energy alignments of the essential Kohn-Sham orbitals near the HOMO-LUMO gap from different functionals, together with their iso-surfaces, are shown for the FM configuration (spin-$\frac{3}{2}$). The energies (in the unit of eV) are referred to that of spin-up HOMO. The sign of the wave function is colour-coded, positive in blue and negative in purple. The essential spin-up orbtials (HOMO-2 to LUMO) and spin-down (LUMO to LUMO+3) are shown. The spin-up orbital HOMO-2 carries the localised spin on Cu and the spin-up orbitals HOMO-1 and HOMO form triplet exciton. Cu is depicted in orange, C in dark gray, and H in light grey. The iso-surface value is set to be $0.02e/\AA^3$.
}\label{fig:orbitals}
\end{figure*}


The Kohn-Sham energy level alignments near the HOMO-LUMO gap for the FM configuration (spin-$\frac{3}{2}$) with the corresponding molecular orbitals are shown in Fig.1. The functionals chosen include further developed PBE functional, namely, PBEh1PBE, hybrid-exchange functionals O3LYP, B3LYP, PBE0, HSE06, and the long-range corrected hybrid-exchange functional CAM-B3LYP.  All the energies (in the unit of eV) are referred to that of the spin-up HOMO (zero-energy). For all these six functionals, the localised orbital carrying Cu spin-$\frac{1}{2}$ is located at the spin-up HOMO-2. This molecular orbital is formed by the hybridisation of $d_{x^2-y^2}$ with ligand $p_x$ and $p_y$ atomic orbitals, having a $b_{1g}$ symmetry that belongs to $\mathrm{D}_{4h}$ point group. The two orbitals for the triplet exciton are located at spin-up HOMO-1 and HOMO. HOMO-1 (HOMO) has a symmetry $a_{1u}$ ($e_{gy}$). This suggests that the triplet exciton is formed by flipping the spin of the HOMO because the HOMO and LUMO in the singlet ground state have $a_{1u}$ and $e_g$ symmetries respectively. This physical picture is consistent with the previous calculation results \cite{marom2008,vas2009}. As shown in Fig.1, the spin-down $a_{1u}$ orbital is unoccupied, which is exactly expected for a formation of the lowest-energy triplet exciton. 

The triplet excitation energy (the energy difference between FM state and singlet ground states) as tabulated in Table.\ref{tb:exchange} ranges from 0.76 eV (B3LYP) to 1.25 (PBE0). This is lower than the HOMO-LUMO gap ($\sim 2 $ eV) that is approximately the energy of the first singlet excited state (mainly dominated by HOMO and LUMO). The gap between the occupied spin-up and unoccupied spin-down $b_{1g}$ orbitals is approximately the on-site Coulomb interaction $U$ of Cu spin). The largest this gap ($\sim 7.4$ eV, see Fig.1) was computed by using the CAM-B3LYP functional. However, the other five functionals, as shown in Fig.1, have given smaller $U$ ($\sim \ 3$ -- $4$ eV, see Fig.1). This might be due to the stronger localization of Cu d-orbital when mixing a larger amount of exact exchange in the long range in CAM-B3LYP.  This gap from the other five functionals is in a good agreement with the previous calculations \cite{wu2011}. The spin-up HOMO-LUMO gaps in this FM configuration are $\sim 1$ eV except that given by CAM-B3LYP ($\sim 3$ eV). The expectation values of $\hat{\vec{S}}^2$ are between $3.77$ and $3.79$, which is close to the expected  ($3.75$) for a spin-$\frac{3}{2}$ object (See Table.\ref{tb:exchange}). 

The electronic structures near the HOMO-LUMO gap for the AFM configuration from different functionals are shown in Fig.2. For all the functionals listed, the spin-down HOMO is the localised orbital derived from $d_{x^2-y^2}$ orbital. The spin-up HOMO ($e_{gy}$) and HOMO-1 ($a_{1u}$) form the triplet exciton on the Pc ring. The expectation value of $\hat{\vec{S}}^2$ is between $1.78$ and $1.80$, which is close to the expected ($1.75$) for a broken-symmetry state of a spin-$1$ and a spin-$\frac{1}{2}$. 

When using the LSDA and GGA exchange-correlation functionals, the desired AFM state was initialised through orbital switching, but after the self-consistent process, the AFM configuration can not be converged. However, in sharp contrast, the AFM configuration can be achieved by using PBEh1PBE, O3LYP, B3LYP, PBE0, HSE06, and CAM-B3LYP functionals. This might suggest that the hybrid-exchange functional may indeed localize the $d$-orbital-derived molecular wave functions, whereas the LSDA and GGA functionals fail doing so. Notice that the tendency of GGA to over-delocalise electrons as compared to hybrid-exchange functionals has been well known \cite{ruiz2004}.

\begin{figure*}[htbp]
\centering
\includegraphics[scale=0.875,clip=true]{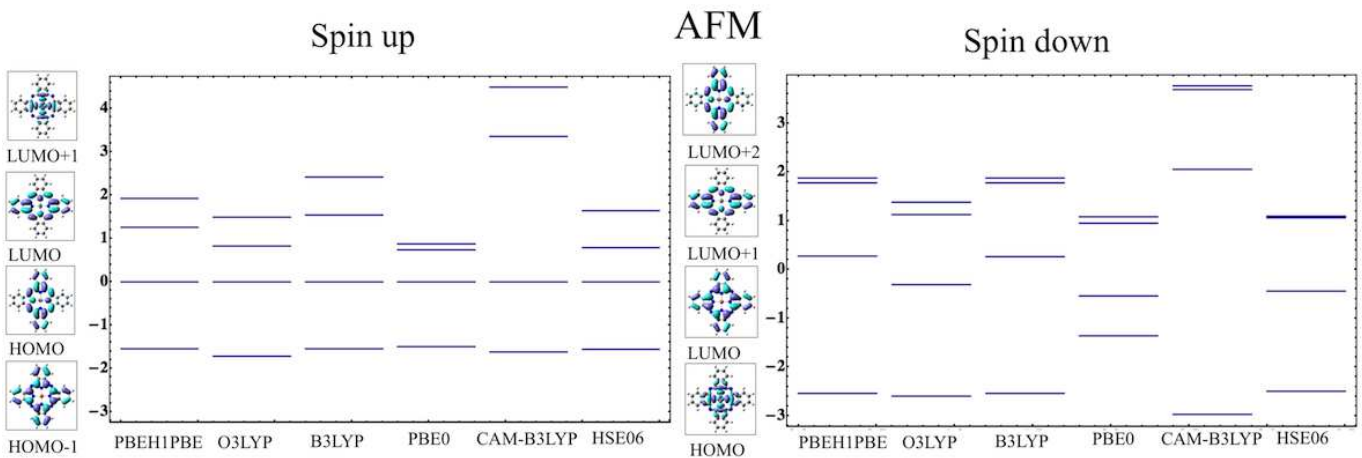}
\caption{(Color online.) The energy alignments of the essential Kohn-Sham orbitals near the HOMO-LUMO gap from different functionals, together with their iso-surfaces, are shown for the AFM configuration (broken-symmetry state). The essential spin-up orbitals (HOMO-1 to LUMO+1) and spin-down (HOMO to LUMO+2) are shown. The energies (in the unit of eV) are referred to that of spin-up HOMO. The spin-down orbital HOMO carries the localised spin on Cu and the spin-up orbitals HOMO-1 and HOMO form triplet exciton. Cu is depicted in orange, C in dark gray, and H in light grey. The iso-surface value is set to be $0.02e/\AA^3$.
}\label{fig:orbitals}
\end{figure*}

\subsection{Exchange interaction}
\begin{table*}[htbp]
    \begin{tabular}{ccccccccccc}
    \hline\hline
      Approximation &S-VWN(LSDA)&PW91&PBE&BLYP&PBEh1PBE&O3LYP&B3LYP&PBE0&cam-B3LYP&HSE06 \\ 
     $E_T$ (eV)&1.24&1.18&1.25 & 1.18 &0.95&1.08&1.25&0.96&0.76&1.14\\
      $\langle\hat{\vec{S}}^2\rangle$ (FM)&3.753&3.756&3.753&3.755&3.777&3.763&3.770&3.776&3.794&3.774\\
      $\langle\hat{\vec{S}}^2\rangle$ (AFM)& - & - & - & - &1.784&1.764&1.776&1.783&1.802&1.781\\
     $J$ (meV)& - & - & - & - &-5.7&-15.9&-9.1&-6.0&-1.5&34.6 \\ \hline
    \end{tabular}
    \caption{The computed triplet-state energies, expectation values of $\hat{\vec{S}}^2$ for FM and AFM (broken-symmetry) configurations, and exchange interactions between Cu spin and ligand triplet $J$ using different approximations to the exchange-correlation functionals are summarized here.}\label{tb:exchange}
\end{table*}
\normalsize

\begin{figure}[htbp]
\includegraphics[scale=0.43,clip=true]{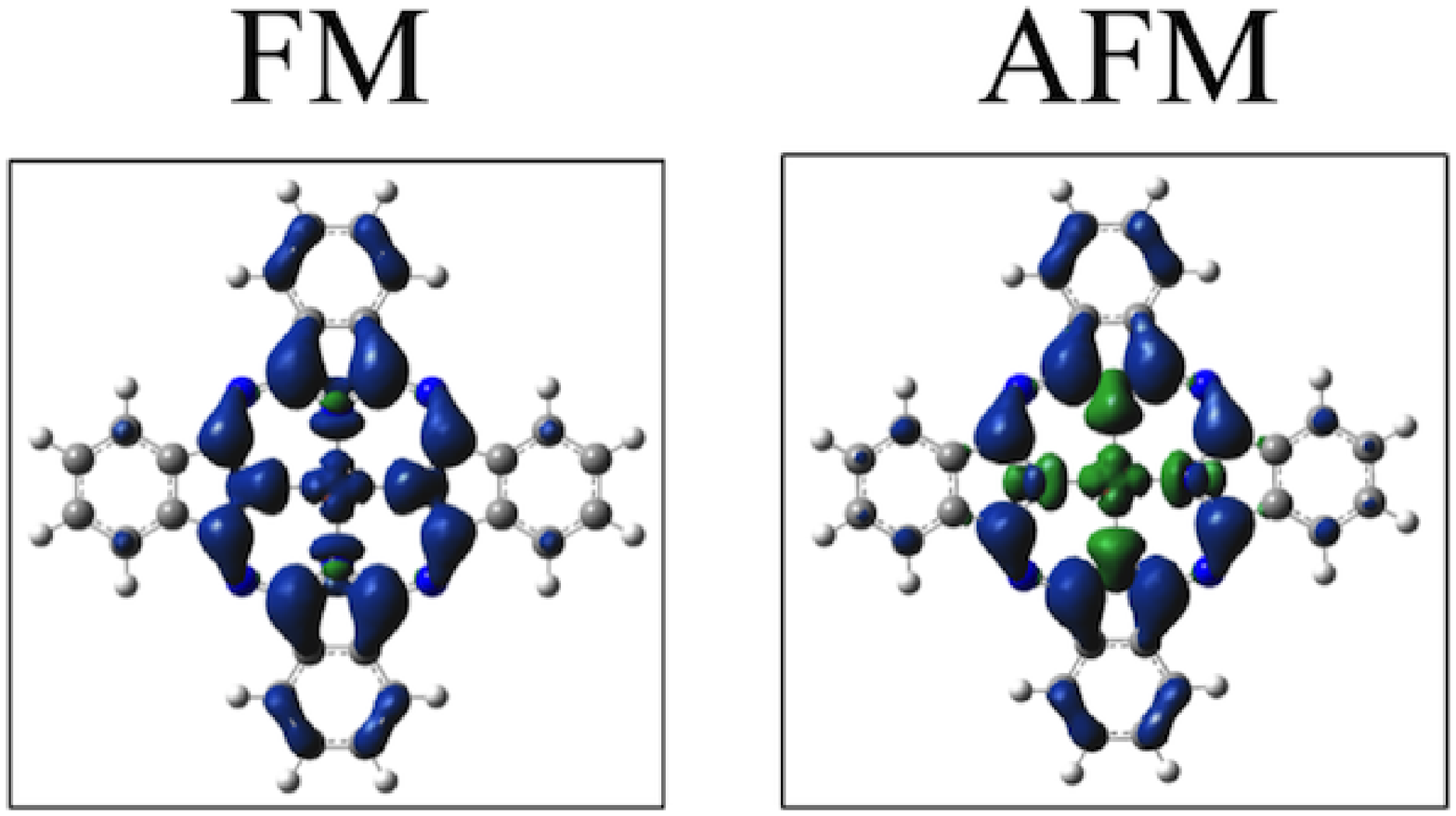}
\caption{(Color online.) The iso-surfaces of spin densities for FM (left) and AFM (right) configurations computed by using B3LYP functional are shown. The positive spin density is depicted in blue and negative in green. Notice that the localised spin is flipped in the AFM configuration as compared to the FM one. The value of the iso-surface is set to be $0.004e/\AA^3$.
}\label{fig:spind}
\end{figure}

The Mulliken spin densities for these two spin-configurations based on B3LYP functional are shown in Fig.\ref{fig:spind}. As expected, the triplet exciton spin density is largely distributed on the Pc ring. The Mulliken spin densities calculated by using the other functionals listed in Fig.1 and Fig.2 share similar qualitative feature as compared with that from B3LYP.  The magnitude of the computed exchange interaction between the triplet and localised spin (Table.\ref{tb:exchange}) is between $1.5$ meV (ferromagnetic, CAM-B3LYP) and $34.6$ meV (anti-ferromagnetic, HSE06). These functionals listed in Fig.1 and Fig.2 have predicted a ferromagnetic interaction except for HSE06 giving an anti-ferromagnetic one. However, these calculated exchange interactions is much larger than the nearest-neighbouring exchange interaction in CuPc crystal (typically $\sim 0.1$ meV) \cite{wu2008,heutz2007} and the hyperfine interaction on CuPc (typically $\sim 10^{-5}$ meV) \cite{marc2012}.

Regarding the exchange mechanism, the super-exchange contribution can be neglected as the orbitals involved in the exchange interaction have different symmetries ($b_{1g}$ for the localised spin; $a_{1u}$ and $e_{g}$ for the triplet exciton). The confinement of a single molecule results in a large overlap between orbitals in the exchange integral, and this would in turn trigger a relatively large ferromagnetic interaction owing to the direct-exchange mechanism. This speculation would therefore lead to a ferromagnetic interaction, which is consistent with the results based on PBEh1PBE, O3LYP, B3LYP, and CAM-B3LYP, but not with HSE06. Supposing the indirect exchange effect is small, then it might be worth comparing the major difference between these functionals. The key feature of HSE06, which distinguishes itself from the other functionals, is that in HSE06 the screened Coulomb potential is used instead of computing exact Fock exchange. This perhaps has a significant effect on the total energies with different spin configurations and hence the exchange interaction. However, more detailed analysis about this sign change of exchange interaction owing to functionals is beyond the scope of this paper, which is expected to be discussed separately. Moreover, the computed exchange interaction is much larger than $\frac{\hbar}{T} (\sim 10^{-4}$ meV if $T \sim 10 \ ns$), which means the spin coherence induced by the exchange interaction could potentially survive the fast relaxation of triplet state in CuPc. Therefore, it is conceivable that this optically induced interaction could be useful for the manipulation of spin-qubit in molecular QIP.

The molecular structure has also been optimised with triplet on the Pc ring. CuPc molecule undergoes a Jahn-Teller distortion with a symmetry reduction from $\mathrm{D}_{4h}$ to $\mathrm{D}_{2h}$ because $e_{gy}$ orbital is singly occupied. The molecule elongates along $y$-direction but shortens along $x$-direction. However, the change of molecular structure is so small (the change of the atomic coordinates is up to 2 \%) that the influence on the exchange interaction is negligible. Nevertheless, the calculations presented here would still be useful when the molecule is still in the ground-state geometry, i.e., before the structural relaxation due to triplet formation happens.

\subsection{Spin manipulation in CuPc by using triplet}

The spin manipulation could be performed either on the CuPc molecule monolayer on the surface or the bulk CuPc crstyal. For the former, the substrates should not be sensitive to microwave field, for example, the non-magnetic semiconducting materials. For the latter, the might be addition modulations from inter-molecular exchange interaction, but this won't be a major problem as its magnitude is much smaller than that between triplet exciton and Cu spin. 

First, a laser pulse can be applied to stimulate the molecule to the singlet excited state followed by the fast relaxation to the triplet ground state.  Then, from this point, the triplet and Cu $\frac{1}{2}$-spin are in a coherent state owning to the exchange interaction. A standard series of electron spin resonance (ESR) pulses can be used to manipulate the spins on CuPc as described in Ref.\cite{fillidou2012}. Notice that the whole ESR pulse sequence should be shorter than the triplet lifetime, i.e., tens of $ns$. For the spin manipulation, a short ESR pulse (pulse length $\sim 0.1 \ ns$) can be implemented first. After $\sim 1 \ ns$ the ESR $\pi$-pulse is applied to refocused the spins in order to obtain a spin echo. By recording the integrated echo, one can observe the time-dependent oscillation of integrated echo, which is essentially an indication of spin rotation, i.e., a process for spin manipulation. The theoretical description of this manipulation process can be derived by using the theory of open quantum system \cite{opensystem}. However, this is beyond the scope of this paper, and will be described in another paper, which would be focused more on the modelling of ESR manipulation of spins in CuPc by using triplet excitation.

\section{Conclusion}\label{conclusions}

The exchange interaction between triplet exciton and localised spin in CuPc has been computed using Dunning's correlation consistent basis set and a variety of functionals within DFT. The computed exchange interactions are in the order of meV. These computed exchange interactions are much larger than the nearest-neighbouring inter-molecular exchange in CuPc, which implies the potential of optical excitation for transient modification of the magnetic properties of CuPc and the application to molecular QIP. This quantity is also much larger than $\hbar/T$, which means quantum operation may be performed well before the triplet state relaxes back down to the singlet ground state. Therefore, in conjugation with the long spin lattice relaxation time ( $\sim \mu s$) of the localised spin in CuPc \cite{marc2012}, this exchange interaction might be useful for the optical manipulation of electron-spin qubit in QIP.

\begin{acknowledgments}

I wish to acknowledge the support of the UK Research Councils Basic Technology Programme under grant EP/F041349/1. I thank Gabriel Aeppli, Andrew Fisher, Nicholas Harrison, Sandrine Heutz, Chris Kay, Marc Warner, and Yunyan Zhang for stimulating discussions.

\end{acknowledgments}

\end{document}